\documentstyle[aps,preprint,epsf]{revtex}

\begin{document}

\draft

\title{Total and Parity-Projected Level Densities of Iron-Region Nuclei
in the Auxiliary Fields Monte Carlo Shell Model}

\author{H. Nakada$^{1,2}$ and Y. Alhassid$^1$}
\address{$^1$ Center for Theoretical Physics,  Sloane Physics Laboratory,
Yale University, New Haven, Connecticut  06520\\
$^2$ Department of Physics, Chiba University,
Yayoi-cho, Chiba 263, Japan}

\date{\today}

\maketitle

\begin{abstract}
 We use the auxiliary-fields Monte Carlo method for the shell model
 in the complete $(pf+0g_{9/2})$-shell to calculate  level densities.
We introduce parity projection techniques which
 enable us  to  calculate  the parity dependence of the level density.
 Results are presented for $^{56}$Fe,
where the calculated total level density is found to be in remarkable agreement
with the experimental level density.  The parity-projected densities are well
 described by a backshifted Bethe formula,  but with significant
dependence of the  single-particle level-density and backshift parameters
 on parity.  We compare our exact results
with those of the thermal Hartree-Fock approximation.
\end{abstract}


 Nuclear level densities are important  for
 theoretical estimates of nuclear reaction rates in nucleosynthesis.
The $s$- and $r$-processes that involve medium-mass and heavier nuclei
are determined by the competition between neutron-capture and $\beta$-decay,
and the neutron-capture cross-sections are strongly affected by the level
 density around the neutron resonance region.
 Reliable estimates of nuclear abundances often require accurate
 level densities.
 For example, the abundance of $s$-process nuclei with non-magic
 neutron number  is (in the local approximation) inversely proportional
to the neutron-capture cross-section \cite{SFC65} which in turn is
proportional to the level density.
Most conventional calculations of the nuclear level density
are based on the Fermi gas model within the grand-canonical ensemble
\cite{ref:BM1}. For a gas of free nucleons one obtains the well-known Bethe
formula.
A simple but useful phenomenological modification is often adopted,
in which the excitation energy  $E_x$ is backshifted
\cite{ref:HWFZ}, giving a total nuclear level density of
\begin{eqnarray}\label{BBF}
 \rho^{\rm BBF} (E_x) =
g  {{\sqrt\pi}\over{24}} a^{-\frac{1}{4}} (E_x - \Delta)^{-\frac{5}{4}}
 e^{2\sqrt{a (E_x - \Delta)}}
\end{eqnarray}
with $g=2$. The backshift $\Delta$ originates in pairing correlations
and shell effects, while
the parameter $a$ is determined by the single-particle level-density
at the Fermi energy. By adjusting the value of  $a$
for each nucleus, the backshifted Bethe formula (BBF)  (\ref{BBF}) fits well
a large volume of experimental data.
The value of the parameter, however, is not well understood;
the Fermi-gas model grossly underestimates the value of $a$,
and cannot account for its exact mass and nucleus dependence.
Consequently, it is difficult to predict the level density to an accuracy
better than an order of magnitude. Much less is known about the
 parity-dependence of the level density.  The finite-temperature mean-field
 approximation\cite{ref:MFA} offers an improvement over the Fermi gas model
but still ignores important two-body correlations, especially
at low temperatures.

In this paper we  study the nuclear level density
in the framework of the interacting shell model,
in which the two-body correlations are fully taken into account
within the model space.
It should be noted, however,  that the finite size of the model space limits
the validity of such calculations to below a certain excitation energy.
The size of the valence shells required
to describe the neutron-resonance region for medium-mass and heavier nuclei
is too large for conventional diagonalization techniques to be practical.
However, the recently proposed
shell model Monte Carlo (SMMC) method\cite{ref:MCSM}
makes it possible to calculate  thermal averages
in much larger model spaces by using fluctuating auxiliary-fields.
As shown below, these methods are particularly suitable for
calculations of level densities.

Nuclei in the iron region play a special role in nucleosynthesis.
They are the heaviest that can be produced
inside normal massive stars,
and  the starting point of the synthesis
of heavier nuclei.
These nuclei  are  in the middle of the $pf$-shell,
and are just beyond the range of nuclei
where conventional shell model techniques can be applied
in a complete $pf$-shell model space\cite{ref:RB,ref:PZ}.   Truncated
shell calculations \cite{ref:NSO}
were successfully used to describe the low-lying states in these nuclei.
However, their neutron separation energy,
typically $E_x\sim 5$--$15$~MeV, is too high to justify such truncation.
The SMMC method was used
to calculate thermal properties of $^{54}$Fe
in a full $pf$-shell\cite{ref:MC-Fe54}
with the Brown-Richter Hamiltonian.
The Monte Carlo sign problem of this realistic interaction
is overcome through the  techniques of Ref. \cite{ref:ADK}.
However, the statistical errors were too large
to obtain accurate level densities.
Furthermore, the energy range of interest in the iron region
($E_x\sim 5$--$15$~MeV) contains negative-parity states
which are not included in the $pf$-shell model space.
In this letter we introduce parity-projection methods for the auxiliary fields,
and use the SMMC  within the full $pf$- and $0g_{9/2}$-shell to calculate
total and parity-projected level densities in the iron region.  This model
space is sufficient to describe both positive- and negative-parity states
for excitation energies up to  $20$~MeV.
To keep the statistical errors small, we construct an interaction
which is free from the Monte Carlo sign problem,
yet  realistic enough to describe  collective features that
 affect the level density.
In particular we present results for  $^{56}$Fe,
for which experimental data are available.

We adopt an isoscalar Hamiltonian of the form \cite{ref:ABDK}
\begin{eqnarray}\label{2}
  H = \sum_a \epsilon_a \hat n_a + g_0 P^{(0,1)\dagger}\cdot \tilde P^{(0,1)}
     - \chi \sum_\lambda k_\lambda O^{(\lambda,0)}\cdot O^{(\lambda,0)} \;,
\end{eqnarray}
where
\begin{eqnarray}\label{3}
 P^{(\lambda,T)\dagger}&=&{\sqrt{4\pi}\over{2(2\lambda +1)}}
 \sum_{ab} \langle j_a\| Y_\lambda \|j_b\rangle
 [a_{j_a}^\dagger \times a_{j_b}^\dagger]^{(\lambda,T)}\;, \nonumber\\
O^{(\lambda,T)}&=&{1\over\sqrt{2\lambda +1}}
 \sum_{ab} \langle j_a\| {{dV}\over{dr}} Y_\lambda \|j_b\rangle
[a_{j_a}^\dagger \times \tilde a_{j_b}]^{(\lambda,T)} \;,
\end{eqnarray}
and $(\cdot)$ denotes a  scalar product in both spin and isospin.
The modified annihilation operator is defined
by $\tilde a_{j,m,m_t} = (-)^{j-m+{1\over 2}-m_t} a_{j,-m,-m_t}$,
and a similar definition is used for $\tilde P^{(\lambda,T)}$.
To conserve the isospin symmetry,
the single-particle energies $\epsilon_a$ are taken to be equal
for protons and neutrons,
and are determined from a Woods-Saxon potential plus spin-orbit interaction
with the parameters quoted in Ref.~\cite{ref:BM1b}.
$V$ in (\ref{3}) is the central part of this single-particle potential.
The multipole interaction in (\ref{3}) is obtained (with $k_\lambda=1$)
by expanding the separable surface-peaked interaction
$v(\bbox{r}, \bbox{r}^\prime)
 = -\chi (dV/dr)(dV/dr^\prime)  \delta(\hat{\bbox{r}} - \hat{\bbox{r}}^\prime)$.
The interaction strength $\chi$ is  fixed by  a self-consistency
relation \cite{ref:ABDK},
and we find $\chi=0.026$~MeV$^{-1}$fm$^2$ for $^{56}$Fe.
In our present calculations we include
the quadrupole, octupole and hexadecupole terms
($\lambda=2,3$ and $4$, respectively).
Since our shell-model configuration space includes the valence shell alone,
core polarization effects are taken into account
by using renormalization factors $k_\lambda$.
We adopt the values $k_2=2$, $k_3=1.5$ and $k_4=1$,
which are consistent with a realistic effective interaction in this shell
derived by the folded-diagram technique\cite{ref:Kuo-pc}.
This interaction satisfies the modified sign rule
(suitable for shells with mixed parities)\cite{ref:ADK},
and therefore has a Monte Carlo sign of $\langle \Phi\rangle=1$
for even-even nuclei. The pairing strength $g_0$ is determined  by using
 the experimental odd-even mass differences\cite{ref:mass}
for nuclei in the mass region $A=40$--$80$ to estimate the pairing gap.
The number-projected BCS calculation is then performed
for fifteen spherical nuclei with $Z=20$, $N=28$, $Z=28$ or $N=40$
to find the value of $g_0$ that will reproduce the estimated pairing gaps.
In contrast to heavier nuclei\cite{ref:Bes},
we find no systematic $A$-dependence in $g_0$,
and a constant mean value of $g_0=0.212$~MeV is adopted.

It is difficult to obtain detailed spectroscopic
information on excited states in the SMMC method.
Instead, it is possible to calculate directly low moments of  strength
functions \cite{ref:ALL}, such as the average energy
of the quadrupole excitation
 $\overline E(Q^{(2,0)}) \equiv
[{\sum_i E_x(2^+_i) \left|\langle 2^+_i |Q^{(2,0)}|0^+_1\rangle \right|^2}]
/[{\sum_i \left|\langle 2^+_i |Q^{(2,0)}|0^+_1\rangle \right|^2}]$.
Here $Q^{(\lambda,T)}$ is defined by replacing $dV/dr$
of $O^{(\lambda,T)}$ in Eq.~(\ref{3}) by $r^\lambda$.
Data on the strength function of the mass quadrupole moment are available
from $(p,p^\prime)$ experiments in a broad energy range
in $^{56}$Fe\cite{ref:NDS56}.
We find  $\overline E_{\rm exp}(Q^{(2,0)}) \simeq 2.16$~MeV,
in fairly good agreement with the theoretical value
of $\overline E_{\rm cal}(Q^{(2,0)}) = 2.12\pm 0.11$~MeV.
While this interaction is not fully realistic
(e.g., no spin degrees of freedom are included),
it seems to reproduce quite well the collective features of the nucleus.
It is thus reasonable to expect that certain gross properties
like the level density are described well by this interaction.  Note,
however, that  possible isospin components of the effective
 interaction that  push up states of higher isospin are not included in
 our present study since they do not have a good Monte Carlo sign.

Since both the $0f_{7/2}$ and $0g_{9/2}$ orbits are included in our model space,
spurious center-of-mass motion may occur.
Although this problem has not been fully explored within the SMMC framework,
it is expected to be unimportant in our case.
 The energy difference between $0f_{7/2}$ and $0g_{9/2}$ is 9.6 MeV,
 so we expect spurious states to appear at excitation energies around
10 MeV and higher. However, their density is comparable to that of the
 non-spurious states but at  about 10 MeV lower in energy, and is thus
 a negligible  fraction of the total density at the actual excitation energy.

 In the SMMC the energy is calculated as a function of  inverse temperature
 $\beta$,  from the canonical expectation value of the Hamiltonian
$E(\beta) \equiv\langle H\rangle_\beta$
through an exact particle-number projection\cite{ref:MC-Nproj}
of both protons and neutrons.
The partition function  $Z(\beta)$ is then determined by a numerical
 integration of $E(\beta)$
\begin{eqnarray}\label{6}
 \ln \left[ Z(\beta)/ Z(0)\right] =
 - \int_0^\beta d\beta^\prime E(\beta^\prime) \;,
\end{eqnarray}
where $Z(0) = {\rm Tr }~\bbox{1}$ is just the total number of states
within the model space.
The level density $\rho(E)$ is the inverse Laplace transform of $Z(\beta)$,
and is calculated in the saddle-point approximation from
\begin{eqnarray}\label{7}
  \rho(E) &=& (2\pi \beta^{-2} C)^{-1/2}  e^{S} \;;
  \nonumber \\
 S(E) & = & \beta E + \ln Z(\beta) \;,
~~\beta^{-2}C(\beta)  = - dE / d\beta \;.
\end{eqnarray}
Here $\beta=\beta(E)$ is determined by inverting the relation $E=E(\beta)$,
and $C$ is the heat capacity calculated
by numerical differentiation of $E(\beta)$.

We now introduce parity-projection techniques in the SMMC
 to calculate parity-projected observables (see Ref.~\cite{Ta81}
 for parity-projected thermal mean-field approximation).
Using the Hubbard-Stratonovich  representation
for $e^{-\beta H}$ and the projection operators
$P_\pm = (1 \pm P)/2$ ($P$ is the parity operator)
on states with positive and negative parity, respectively, we can  write
the projected energies  $E_\pm (\beta) \equiv { {\rm Tr} (H P_\pm e^{-\beta H} )
 / {\rm Tr} (P_\pm e^{-\beta H})}$ in the form

\begin{eqnarray} \label{E_pm}
 E_\pm (\beta)  = {\int D[\sigma]W(\sigma)  \left[ \langle H \rangle_\sigma \pm
\langle H \rangle_{P\sigma} \zeta_P(\sigma) / \zeta(\sigma) \right]
\over
\int D[\sigma]W(\sigma) \left[  1 \pm {\zeta_P(\sigma) /
\zeta(\sigma)} \right] }
\;.
\end{eqnarray}
The integration over the auxiliary fields $\sigma$
is performed with  the usual  Monte Carlo weight function
$W(\sigma) = G(\sigma) \zeta(\sigma)$,  where
$G$ is a gaussian factor  and $\zeta(\sigma) = {\rm Tr}~U_\sigma$ is the
 partition function  of the non-interacting
propagator $U_\sigma$ \cite{ref:MCSM}.  In (\ref{E_pm})
 $\zeta_P(\sigma) \equiv {\rm Tr}(PU_\sigma)$ and
$\langle H \rangle_{P\sigma} \equiv {\rm Tr}(HPU_\sigma)/{\rm Tr}(PU_\sigma)$.
Since the parity operator can be expressed
as a product of the corresponding parity operators of each of the particles,
it follows that the operator $PU_\sigma$ can be represented
in the single-particle space by the matrix ${\cal P} {\cal U}_\sigma$,
where ${\cal U}_\sigma$ is the matrix representing $U_\sigma$
 in the single-particle space,
and ${\cal P}$ is a diagonal matrix with elements $(-)^{\ell_i}$
($\ell_i$ is the orbital angular momentum of the single-particle orbit $i$).
This representation allows the calculation of $\zeta_P(\sigma)$ and
$\langle H \rangle_{P\sigma}$  through matrix algebra in the
single-particle space, similar to the calculation of  $\zeta(\sigma)$ and
$\langle H \rangle_\sigma$, except that the matrix ${\cal U}_\sigma$
is replaced by ${\cal P} {\cal U}_\sigma$.
Once we calculate $E_\pm(\beta)$, we can proceed to calculate
the densities $\rho_\pm(E)$ as for the total level density.

\begin{figure}

\vspace{ 2 cm}

\centerline{\epsffile{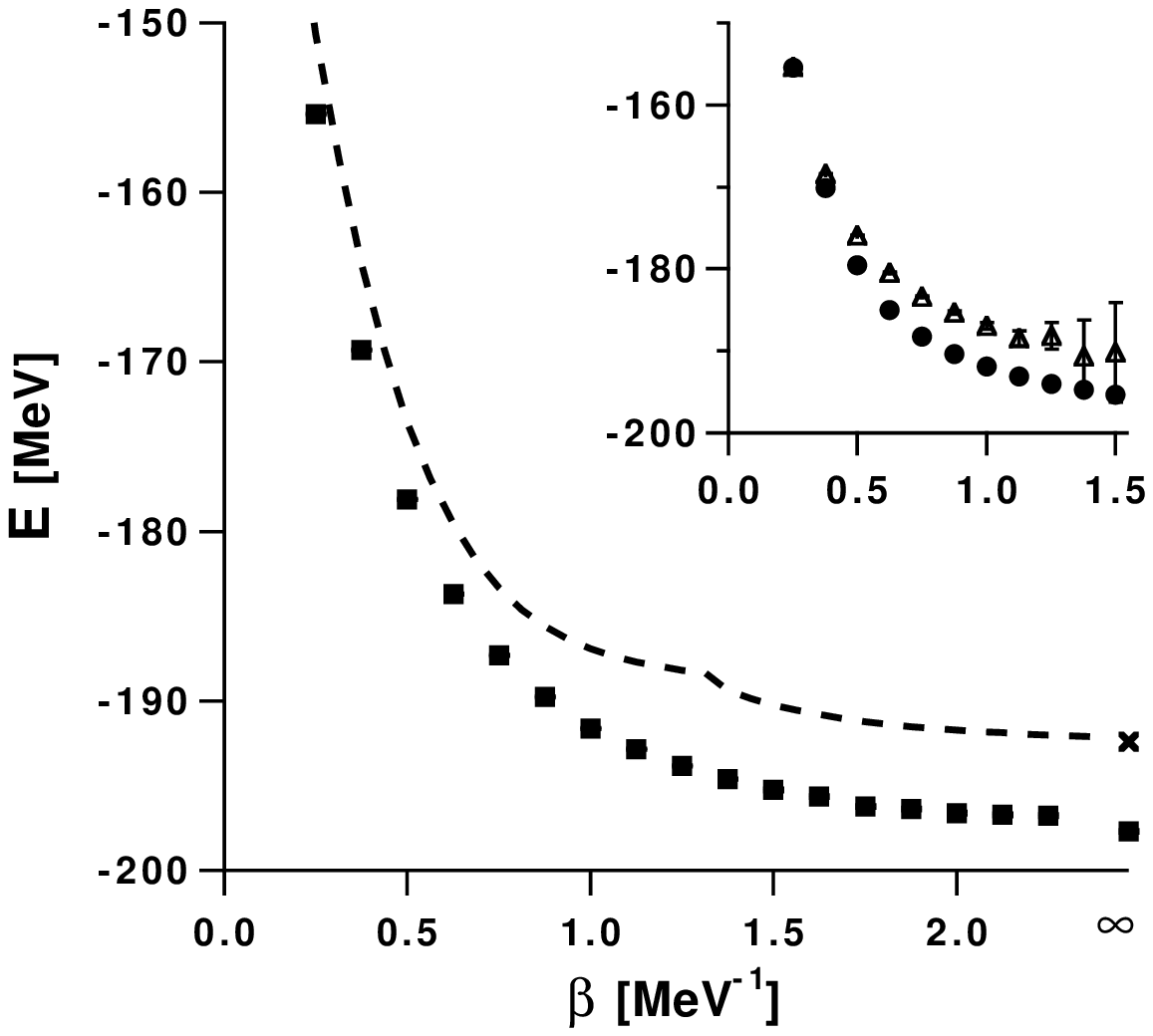}}

\vspace{2 cm}

\caption{ The total energy $E$ as a function of $\beta$ for $^{56}$Fe.
The SMMC values are shown by solid squares,
while the HFA values by a dashed line.
The inset shows  the positive (circles) and  negative (triangles)
parity-projected  SMMC energies. }
\label{fig:E(beta)}

\newpage

\vspace*{2 cm}

\centerline{\epsffile{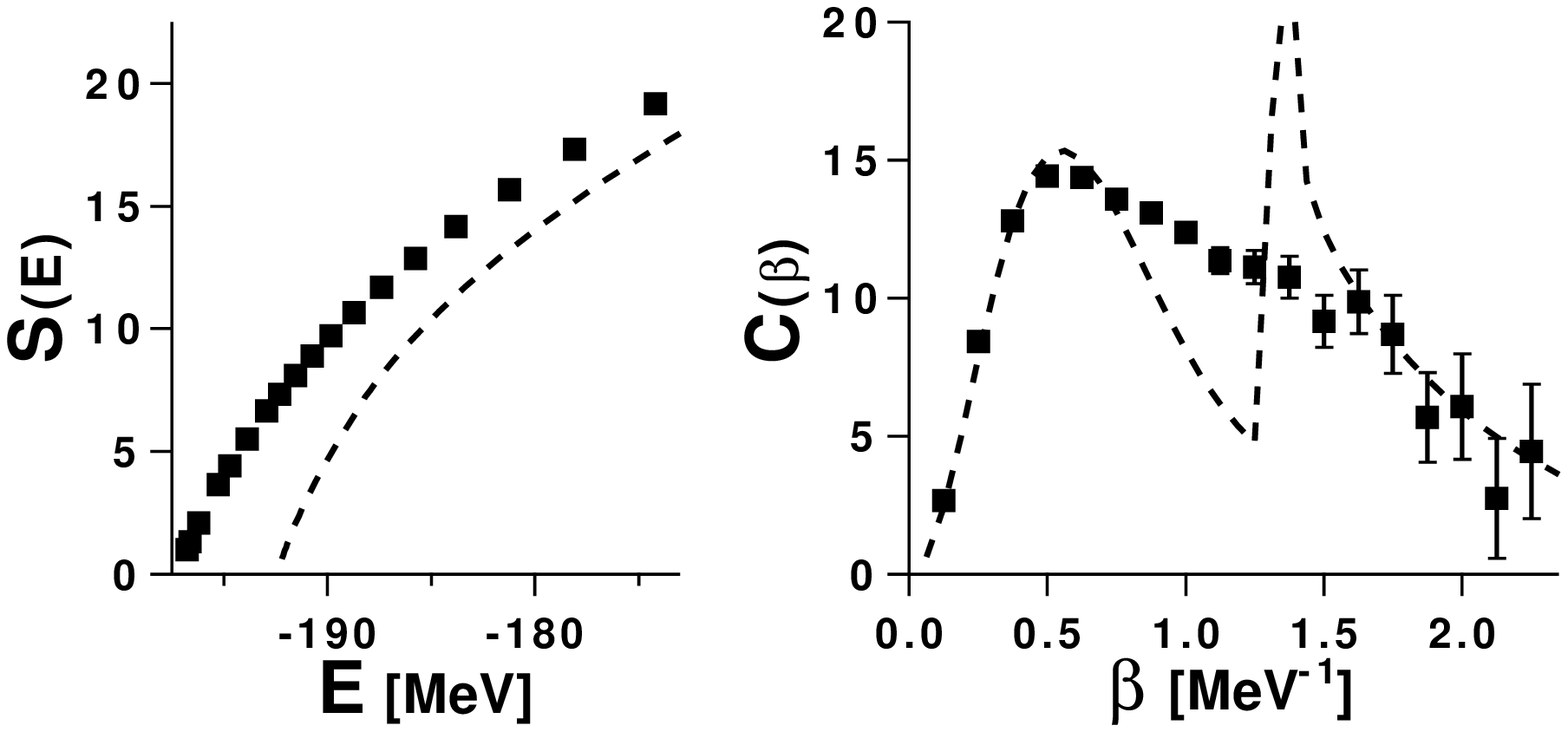}}

\vspace{2 cm}

\caption{Right: the entropy as a function of  energy $E$.
Left:  the heat capacity $C$ as a function of $\beta$.
The conventions are as in Fig.~1. }
\label{fig:C(T)}

\newpage

\vspace*{2 cm}

\centerline{\epsffile{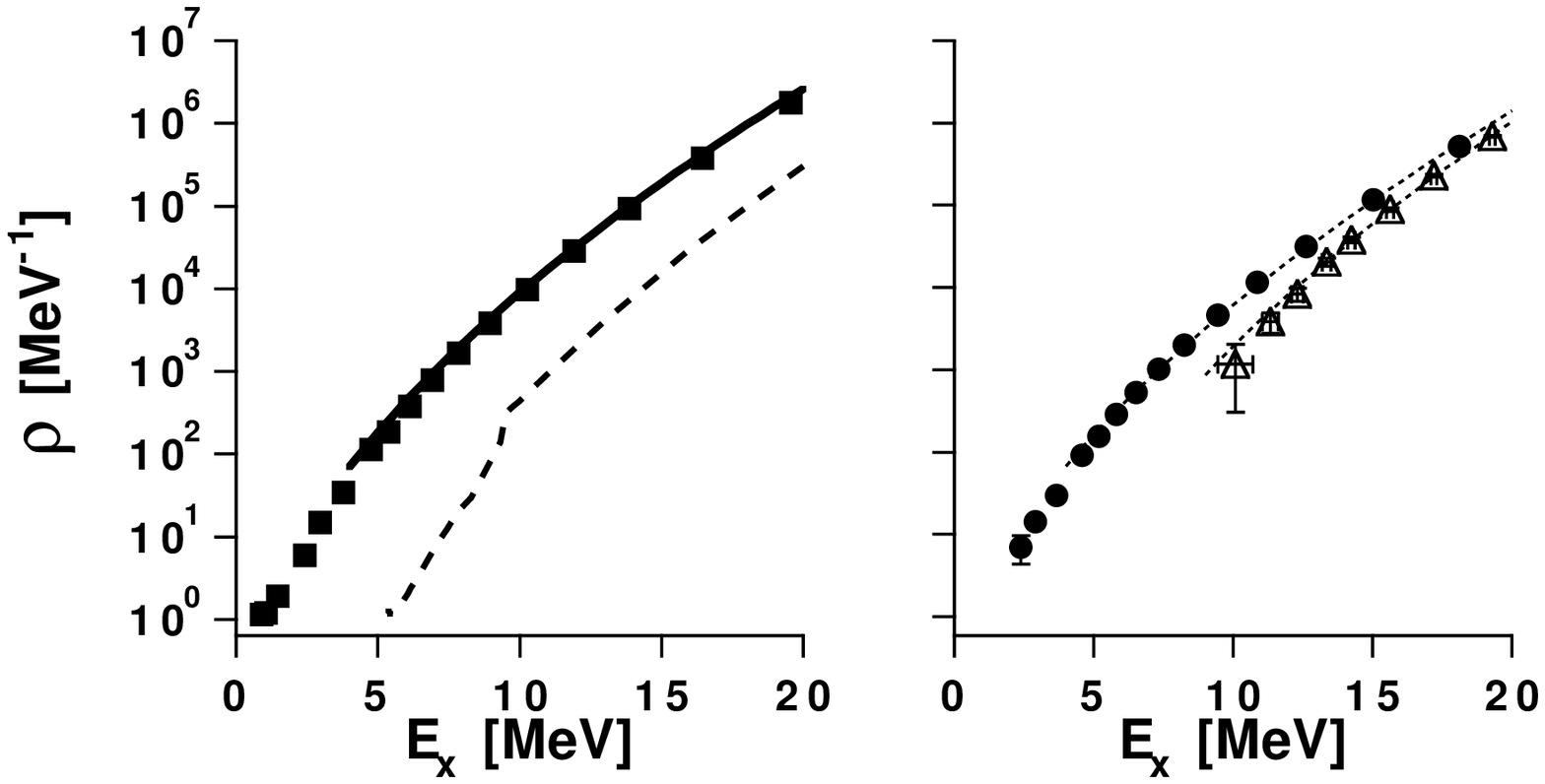}}

\vspace{2 cm}

\caption{Level densities of $^{56}$Fe.
Left: total level density.
The SMMC level density  (solid squares) is
 compared with the HFA level density (dashed line).
The  solid line is the experimental  level density \protect\cite{ref:WFHZ}.
Right: positive- and negative-parity  level densities in the SMMC.
The conventions are as in the inset to Fig.~1.  The dotted lines are the fit to
Eq.~(\protect\ref{BBF}) with the parameters quoted in the text.
}
\label{fig:rho}

\end{figure}

In the following we present results for $^{56}$Fe.
For each  $\beta$  we used a Monte Carlo time slice
of $\Delta\beta=0.03125$~MeV$^{-1}$, and collected more than 4,000 samples.
To describe the level density as a function of the excitation energy  $E_x$,
we also need to know the ground-state energy.
The latter is calculated by extrapolating $E(\beta)$ to
$\beta\rightarrow\infty$.
Fig.~\ref{fig:E(beta)}  shows the energy $E$ as a function of $\beta$.
The SMMC results are the solid squares and include statistical errors
(although the errors are often too small to be visible in the figure).
We compare our results with those of
the thermal Hartree-Fock approximation (HFA)\cite{ref:MFA},
where we observe  large deviations
 at low temperatures. In the HFA, a shape phase-transition occurs
around $\beta \sim 1.3$ MeV$^{-1}$ from a spherical configuration
 (at higher temperatures)
to a deformed one, and its  signature  is observed in the bending of $E(\beta)$.
The inset to Fig.~\ref{fig:E(beta)} presents $E_\pm$ as a function of $\beta$,
calculated using the parity-projection technique of Eq.~(\ref{E_pm}).
Because of the energy gap between $pf$ and $0g_{9/2}$,
$E_-(\beta)$ is notably higher than $E_+(\beta)$ at low temperatures.

The SMMC and HFA  entropies $S$  (versus $E$) and
 heat capacities $C$ (versus $\beta$) are shown
 in Fig.~\ref{fig:C(T)}.
We observe a substantial enhancement in the SMMC entropy over
the HFA entropy, which is due to the full inclusion of the two-body
correlations.
The heat capacity is useful for determining the range of excitation energies
 for which the present model space is sufficient.
The decrease of $C$  at high excitation energy (i.e., small $\beta$)
is associated  with the truncation of the single-particle space,
and we conclude that our present  calculation is meaningful
up to $E_x\sim 20$~MeV.
The discontinuity of the heat capacity at $\beta\sim 1.3$~MeV$^{-1}$ in the HFA
is a signature of the shape transition, but this effect
is washed out in the SMMC.

The SMMC  total level density is shown
in Fig.~\ref{fig:rho} (left panel) as a function of $E_x$.
Although it is difficult to measure the total level density directly,
it can be reconstructed from a few parameters
that are determined experimentally.
The solid line in Fig.~\ref{fig:rho} shows this experimental level density
with the BBF parameters of $a=5.80$~MeV$^{-1}$
and $\Delta=1.38$~MeV \cite{ref:WFHZ}.
Our SMMC result is in excellent agreement
with the experimental level density.   The  slight discrepancy at low energies
may be ascribed to deviations of the moment-of-inertia parameter
\cite{ref:Dilg}
 from its rigid-body value: a rigid-body moment was assumed in deriving the
 the experimental values of $a$ and $\Delta$.
We can also use our microscopically calculated level densities
to extract the level density parameters via a fit to Eq.~(\ref{BBF}).
Using the energy range $4.5~{\rm MeV}<E_x<20~{\rm MeV}$,
we obtain $a = 5.780\pm 0.055~{\rm MeV}^{-1}$
 and $\Delta = 1.560\pm 0.161~{\rm MeV}$.
We note that the statistical errors of the present calculations
are substantially smaller than those of  Ref.\cite{ref:MC-Fe54},
where thermal properties of $^{54}$Fe were calculated
using a realistic interaction in a smaller configuration space ($pf$-shell).
Consequently, accurate level densities can be calculated in the present work.
Also shown in Fig.~\ref{fig:rho} is the level density in the HFA, where
 the excitation energy has been corrected by the difference between
 the mean-field and SMMC ground state energies.
The SMMC level density is significantly enhanced in comparison with
the HFA level density.
The kink around  9~MeV in the HFA level density is  related
to the shape transition, and disappears in the SMMC. Although the 
Hartree-Fock-Bogoliubov approximation is expected to  improve  the HFA
because of the $T=1$ pairing component, the shortcomings of the mean-field
approximation are likely to remain. Indeed,  even when the pairing
force  is omitted from the interaction, we still find that the HFA level density
has a kink and deviates strongly from the exact SMMC level density.

 The Fermi gas model  predicts
 equal positive- and negative-parity level densities at all energies.
However,  this is unrealistic in the neutron resonance regime where
 the neutron resonance energy is comparable to
or even smaller than the energy gap among major shells.
The SMMC results for the parity-projected level densities of $^{56}$Fe
are shown in the right panel of Fig.~\ref{fig:rho}.
They can be well-fitted to a BBF (\ref{BBF}) with $g=1$,  but with
 parity-specific parameters $a_\pm$ and $\Delta_\pm$.
We find $a_+ = 5.611\pm 0.073 ~{\rm MeV}^{-1},~
 \Delta_+ = 0.550\pm 0.196 ~{\rm MeV}$ and
$a_- = 6.209\pm 0.625 ~{\rm MeV}^{-1},~
 \Delta_- = 3.172\pm 1.637 ~{\rm MeV}$.
We remark that negative-parity states in $^{56}$Fe are possible
only when the $g_{9/2}$ level is populated.
Because of  the energy gap between the $pf$ and $g_{9/2}$ orbits
we expect the negative-parity level density to be lower
than the positive-parity density at low energies.
Thus the backshift $\Delta_-$ should be larger than $\Delta_+$,
in agreement with our results.
Both  $\Delta_\pm$ are significantly different
from  $\Delta$ of the total level density.
On the other hand, $a_+$ is rather close to $a$,
while $a_-$ is larger than $a_+$ (and $a$).
At high excitation energies  the Fermi-gas model is
expected to be a reasonable approximation, implying the approximate equality
of  positive- and negative-parity level densities (in our case
$\rho_+\simeq\rho_-$ above  $E_x\simeq  17$~MeV).
Therefore, in the low energy region the negative-parity density is
 expected to rise more quickly
as a function of energy, i.e., $a_- > a_+$.
This relation is confirmed by the present calculations.
So far there has been no systematic study of the parity-dependence
of the level density parameters. It would be interesting to investigate
how this parity-dependence affects the neutron-capture reaction rates.

In conclusion, we have used the auxiliary-field Monte Carlo methods
to calculate the level density of $^{56}$Fe
in the complete $pf$- and $g_{9/2}$-shell,
and found remarkable agreement with the experimental level density.
 The SMMC calculations are an important improvement over the finite
 temperature Hartree-Fock approximation.
We have  introduced  a novel parity-projection technique in the SMMC,
which allows us to study the parity dependence of
both the single-particle level density parameter $a$
and the backshift parameter.
Work in progress includes a systematic study of level densities for nuclei
in the $(pf+g_{9/2})$-shell.

This work was supported in part by the Department of Energy grant
No.\ DE-FG-0291-ER-40608,
and by the Ministry of Education, Science and Culture of Japan
(Grant-in-Aid for Encouragement of Young Scientists, No.\ 08740190).
Y.A.\ thanks G.F. Bertsch  for useful discussions.
H.N.\  thanks F. Iachello for his hospitality at Yale
University, and the Nishina Memorial Foundation for support during his stay.
We acknowledge S. Lenz for his computational help.
Computational cycles were provided by Fujitsu VPP500
at the RIKEN supercomputing facility,
and  we  acknowledge the assistance of  I. Tanihata and S. Ohta.

\end{document}